\documentclass[12pt]{article}
\usepackage{latexsym}
\usepackage{amssymb}
\usepackage{cite}
\input epsf
\newcommand{\sect}[1]{\setcounter{equation}{0}\section{#1}}

\begin{document}

\title{Quantization of Four-form Fluxes\\ and Dynamical
Neutralization\\ of the Cosmological Constant}

\author{
{\sc Raphael Bousso}\thanks{\it
    bousso@hbar.stanford.edu} \\[.3 ex]
    {\it Department of Physics, Stanford University,}\\
    {\it Stanford, California 94305-4060}
\\[1.4ex]
{\sc Joseph Polchinski}\thanks{\it joep@itp.ucsb.edu}
\\[.3 ex] {\it Institute for Theoretical Physics, University of
       California,}\\
  {\it Santa Barbara, California 93106-4030}
}

\date{SU-ITP-00-12~~~NSF-ITP-00-40~~~hep-th/0004134}

\maketitle

\begin{abstract}

A four-form gauge flux makes a variable contribution to the
cosmological constant.  This has often been assumed to take continuous
values, but we argue that it has a generalized Dirac quantization
condition.  For a single flux the steps are much larger than the
observational limit, but we show that with multiple fluxes the allowed
values can form a sufficiently dense `discretuum'.  Multiple fluxes
generally arise in M theory compactifications on manifolds with
non-trivial three-cycles.  In theories with large extra dimensions a
few four-forms suffice; otherwise of order 100 are needed.  Starting
from generic initial conditions, the repeated nucleation of membranes
dynamically generates regions with $\lambda$ in the observational
range.  Entropy and density perturbations can be produced.

\end{abstract}

\pagebreak

\sect{Introduction}
\label{sec-intro}

The cosmological constant problem is one of the central challenges in
quantum gravity, and it continues to provoke novel and interesting
ideas.\footnote{For a classic survey of various approaches see
Weinberg's review~\cite{Wei89}.  A brief survey of recent ideas is
given in Ref.~\cite{Wit00}.}  In some approaches, the cosmological
`constant' becomes a dynamical variable: it is locally constant but
has a continuous range of allowed values, with the effective value in
our universe determined by some dynamical principle.

There are several mechanisms by which the cosmological constant can
become a dynamical variable.  One is through the existence of a
four-form field strength~\cite{AurNic80,DufNie80,Haw84b}.  The
equation of motion requires that such a field strength be constant, so
it has no local dynamics but contributes a positive energy density,
which can cancel a cosmological constant coming from other sourcs if
the latter is negative.  A second mechanism is fluctuations of the
topology of spacetime (wormholes).  Under a plausible interpretation
of the path integral for quantum gravity these convert all constants
of Nature into dynamical variables~\cite{Col88,GidStr88}, though there
is serious doubt as to whether this effect exists in a real theory of
quantum gravity.  A third mechanism is the existence of naked
singularities in compactified dimensions~\cite{Ark00,Kac00}, where the
undetermined boundary conditions at the singularity become a variation
of the effective four dimensional Lagrangian.

In this paper we will discuss certain aspects of the four-form idea,
though at the end we will note that very similar considerations may
apply to the naked singularity.  Our first point is that the four-form
field strength, although usually assumed to take continuous values, is
in fact quantized.  This quantization might be evaded in a purely
four-dimensional theory, but is certainly necessary when gravity is
embedded in a higher-dimensional theory such as M theory.  The size of
the quantum is fixed by microscopic physics, and so the spacing of
energy densities is enormous compared to the actual value, or bound,
on the cosmological constant.  Therefore the four-form cannot play the
assumed role of producing a small cosmological constant.

This is discouraging, but there is a variant of the four-form idea
which has some interesting features.  Typical M theory
compactifications have extra four-form field strengths, arising from
nontrivial three-cycles in the compactification.  If there are several
four-form field strengths, with incommensurate charges, then the
allowed cosmological constants may form a closely spaced `discretuum',
with one or more values in the experimentally allowed range.  The
universe can reach such a value, starting from a larger density,
through a series of domain wall nucleations.  This resembles an idea
of Brown and Teitelboim~\cite{BroTei87,BroTei88}, but has some unique
and attractive features.  In particular, there is a plausible
mechanism for heating the universe after nucleation produces a small
cosmological constant.

A second complication is that in higher dimensional theories there are
in general moduli, and the four-form does not produce a constant
energy density but rather a potential for the moduli.  The analysis of
the four-form flux therefore cannot be separated from the
consideration of the stability of the compactification.  This is a
difficult issue for a number of reasons, and aside from a brief
discussion will sidestep it by working in an artificial, model where
the charges are frozen in incommensurate ratios.  Although this is
rather optimistic, it may be that some features of the cosmology that
we find will survive in more realistic circumstances.

In Sec.~\ref{sec-fourform} we review the physics of four-form fluxes
and explain how a generalized Dirac quantization condition constrains
the value of the four-form flux.  We then investigate the level
spacing in a theory of many four-forms, as generally arise in M theory
compactifications.  We find that the discretuum is sufficiently dense
if there is a membrane charge of order $10^{-1}$ and a large number of
fluxes, say 100.  The large dimension scenario~\cite{ArkDim98}
produces much smaller charges, and can lead to a sufficiently dense
discretuum for as few as four fluxes.

In Sec.~\ref{sec-cosmology} we discuss the resulting cosmology.  We
review the Brown-Teitelboim scenario, in which a cosmological constant
is neutralized by nucleation of membranes.  We then extend this to
multiple four-forms.  If the flux density is initially large, so that
the cosmological constant is positive, then one obtains a picture much
like eternal inflation, where the cosmological constant takes
different values in different expanding bubbles.  De Sitter thermal
effects provide a natural solution to one of the serious problems of
the Brown-Teitelboim idea.  The inflaton can be stabilized in the
inflationary part of its potential until the nucleation reduces the
cosmological constant to near zero, at which point it begins to roll.
This is possible because with multiple four-forms the individual jumps
in the cosmological constant can be quite large.  In the end the
observed cosmological constant is small for anthropic reasons, but in
the weakest sense: we have a universe with different cosmological
constants in different regions, and with galaxies only in regions of
small cosmological constant.  In many respects our picture resembles
an idea of Banks~\cite{Ban86}.  Another example of a discretuum is the
irrational axion~\cite{BarEli83,BDS91}.  While this work was being
completed we learned that Feng, March-Russell, Sethi, and Wilczek are
also considering extensions of the mechanism of Brown and Teitelboim.

\sect{Four-form quantization}
\label{sec-fourform}

\subsection{Four-form energetics}

We first review the basic physics of four-form field strengths.  For
antisymmetric tensor fields, the language of forms is used when
convenient; this is indicated by bold face ($\mathbf{F_{4}}$).  Normal
fonts are used for index notation, or when index notation is implied,
e.g.\ $F_4^2 = F_{\mu\nu\rho\sigma} F^{\mu\nu\rho\sigma}$.

The action for gravity with a bare vacuum energy $\lambda_{\rm bare}$
plus four-form kinetic term is
\begin{equation}
S = \int d^4\!x\, \sqrt{-g} \biggl( \frac{1}{2\kappa_4^2} R -
    \lambda_{\rm bare} - \frac{Z}{2\cdot4!} F_{4}^2 \biggr)
+ S_{\rm branes}\ ,
\end{equation}
where $\mathbf{F_4} = \mathbf{dA_3}$.  We include a general
normalization constant $Z$ in the kinetic term for later convenience.
Certain boundary terms must be added to this action.  They do not
affect the equations of motion and will not be prominent in the
remainder of this paper.  However, they are crucial for the correct
evaluation of the on-shell action when physical quantities are
measured on an equal time hypersurface $\Sigma$.  The usual
Gibbons-Hawking term~\cite{GibHaw77b} is given by
\begin{equation}
S_{\rm GH} =   \frac{1}{\kappa_4^2} \int_{\Sigma} d^3\!x\,
        \sqrt{h} K\ .
\end{equation}
For the four-form field the following boundary term must be included
to obtain stationary action under variations that leave $F$ fixed on
the boundary~\cite{DunJen90}:
\begin{equation}
S_{\rm DJ} =  \frac{Z}{3!} \int d^4\!x\, \partial_\mu
        \left( \sqrt{-g}
        F^{\mu\nu\rho\lambda} A_{\nu\rho\lambda} \right) \ .
\label{eq-bt}
\end{equation}
On shell its value is negative twice the $F^2$ contribution in the
volume term of the action.  This removes the apparent
discrepancy~\cite{Duf89} between the cosmological constant in the
on-shell action and in the equations of motion.

Ignoring the brane sources (we will consider them shortly), the
four-form equation of motion is $\partial_\mu \left( \sqrt{-g}\,
F^{\mu\nu\rho\sigma} \right) = 0$, with solution
\begin{equation}
F^{\mu\nu\rho\sigma} = c
\epsilon^{\mu\nu\rho\sigma}\ ,
\end{equation}
where $\epsilon^{\mu\nu\rho\sigma}$ is the totally antisymmetric
tensor and $c$ is any constant.  Thus there is no local dynamics.  One
has $F_4^2 = -24 c^2$, and so the on-shell effect of the four-form is
indistinguishable from a cosmological constant term.  The Hamiltonian
density is given by
\begin{equation}
\lambda = \lambda_{\rm bare} - \frac{Z}{48} F_4^2 =
\lambda_{\rm bare} + \frac{Zc^2}{2}
\ .
\label{eq-grid-1D}
\end{equation}

Only $\lambda$ is observable: $\lambda_{\rm bare}$ and the four-form
cannot be observed separately in the four-dimensional theory.
Therefore, the bare cosmological constant can be quite large.  For
example, it might be on the Planck scale or on the supersymmetry
breaking scale.  In order to explain the observed value of the
cosmological constant, $\lambda_{\rm bare}$ must be very nearly
cancelled by the four-form contribution.

\subsection{Four-form quantization}

In the original work~\cite{Haw84b}, and in many recent applications,
it as assumed that the constant $c$ can take any real value, thus
cancelling the bare cosmological constant to arbitrary accuracy.
However, we are asserting that the value of $c$ is quantized.  Since
this is somewhat counterintuitive, let us first discuss two things
that the reader might think we are saying, but are not.

First, if there is a gravitational instanton, a Euclidean
four-manifold $X$, then it is natural to expect that the integral of
the Euclidean four-form over $X$ is quantized,
\begin{equation}
\int_X \mathbf{F_{4}} = \frac{2\pi n}{e}\ ,
\quad n \in {\bf Z} \ . \label{eq-tei}
\end{equation}
This is the generalized Dirac quantization
condition~\cite{Sav77,Orl82,Nep85,Tei86}.  It arises from considering
the quantum mechanics of membranes, which are the natural objects to
couple to the potential $\mathbf{A_{3}}$,
\begin{equation}
S = e \int_W \mathbf{A_{3}}
\end{equation}
with $e$ the membrane's charge and $W$ its world-volume.  The
condition that membrane amplitudes be single-valued then implies the
quantization~(\ref{eq-tei}).  This is true, but we are asserting
something in addition: that the actual {\em value} of $\mathbf{F_{4}}$
(or, more precisely, $c$) is quantized, in addition to the integral.

Of course, the inclusion of membranes means that $c$ is no longer
globally constant, as the membranes are sources for $\mathbf{F_4}$.
The value of $c$ jumps across a membrane,
\begin{equation}
\Delta c = \frac{e}{Z} \ .
\end{equation}
The total change in $c$ due to nucleation of any number of membranes
is then a multiple of $e/Z$.

However, it is not this change that we are asserting is quantized, but
the actual value:
\begin{equation}
c = \frac{en}{Z} \ ,\quad n \in {\bf Z}\ . \label{eq-cquant}
\end{equation}
This may seem surprising, but in fact is quite natural.  String theory
has the satisfying property that for every gauge field there exist
both electric and magnetic sources.  This implies a quantization
condition both for the field strength and its dual.  The dual of a
four-form is a zero-form,
\begin{equation}
*\mathbf{F_{4}} = \mathbf{F_{0}}\ .
\end{equation}
A zero-form is naturally integrated over a zero-dimensional manifold,
which is to say that it is evaluated at a point.  The generalized
Dirac condition is that this be quantized, which is precisely
Eq.~(\ref{eq-cquant}):
\begin{equation}
\mathbf{F_{0}} = \frac{en}{Z}\ ,\quad n \in {\bf Z}\ .
\label{eq-teizero}
\end{equation}
The quantizations~(\ref{eq-tei}) and~(\ref{eq-teizero}) are in just
the usual relation~\cite{Tei86} for $n$-form and $(d-n)$-form field
strengths in $d$ spacetime dimensions.

Although natural, it is not clear that the quantization of
$\mathbf{F_{0}}$ is necessary.  The quantization of $\mathbf{F_{4}}$
arises from the consistency of the quantum mechanics of 2-branes, but
that of $\mathbf{F_{0}}$ would come from the quantum mechanics of
$(-2)$-branes, and it is not clear what this should be.  Further,
there is the example of the Schwinger model, where the non-integer
part of $\mathbf{F_{0}}$ is just the $\theta$-parameter, which can
take any real value.

Nevertheless, the quantization condition~(\ref{eq-teizero}) is
necessary when the four-dimensional theory is embedded in string
theory.\footnote {This observation grew out of Ref.~\cite{PolStr96},
where quantization of a top-form (or zero-form) field strength first
appeared.}  Consider for example the compactification of M theory on a
seven-manifold $K$.  We begin with the eleven-dimensional action
\begin{equation}
S = {2\pi M_{11}^9} \int d^{11}\!X\, \sqrt{-g_{11}} \left( R -
    \frac{1}{2 \cdot 4!}  F_4^2 \right) + S_{\rm
    branes}\ ,
\label{eq-act1}
\end{equation}
where we omit the Chern-Simons and fermion terms, which will play no
role.  With this normalization the M2-brane tension and charge are
$2\pi M_{11}^3$, and the M5-brane charge and tension are $ 2\pi
M_{11}^6$.\footnote {For a review see Ref.~\cite{DufKhu94}.} The
M5-brane couples to $\mathbf{A_{6}}$,
\begin{equation}
{2\pi M_{11}^6} \int_{W} \mathbf{A_{6}}\ ,
\end{equation}
where $W$ is the M5 world-volume, and
\begin{equation}
\mathbf{dA_{6}} = \mathbf{F_{7}} = *_{11}\mathbf{F_{4}}\ ,
\end{equation}
where a subscript is used to distinguish the dual in eleven-dimensions
from that in four dimensions.  By the generalized Dirac quantization
it follows that
\begin{equation}
{2\pi M_{11}^6} \int_K \mathbf{F_{7}} = 2\pi n\ , \quad n \in {\bf
Z}\ .
\label{eq-quant7}
\end{equation}

Now reduce to four dimensions.  The eleven dimensional
$\mathbf{F_{4}}$ reduces directly to a four dimensional
$\mathbf{F_{4}}$, with action
\begin{equation}
S = V_7 {2\pi M_{11}^9} \int d^{4}\!x\, \sqrt{-g} \left(
R - \frac{1}{2 \cdot 4!}  F_4^2 \right) + S_{\rm
    branes}\ ,
\label{eq-act4}
\end{equation}
where $V_7$ is the volume of $K$.  Further, the
condition~(\ref{eq-quant7}) becomes
\begin{equation}
\mathbf{F_{0}} = \frac{ n}{M_{11}^6 V_7}\ , \quad n \in {\bf Z}\ .
\label{eq-quant4}
\end{equation}
That is,
\begin{equation}
(2\kappa_4^2)^{-1} = Z = {2\pi M_{11}^9  V_7} \ .
\label{eq-red}
\end{equation}
The quantization~(\ref{eq-quant4}) matches that found in
Eq.~(\ref{eq-teizero}) with $e = 2\pi M_{11}^3$, which is just the
M2-brane tension.

\subsection{Discussion}

The quantization that we have found rules out the precise cancellation
of the cosmological constant that has been assumed in many
discussions.  Brown and Teitelboim~\cite{BroTei87,BroTei88} considered
the approximate neutralization of the cosmological constant by a field
strength taking discrete values (see also Abbott~\cite{Abb85} for a
closely related idea).  In order that this be natural, the spacing
between allowed values of $\lambda$ must be of order the observational
bound.  Since $d\lambda/dn = 2n e^2/Z$ and $n_{\rm final} \approx
\sqrt{|\lambda_{\rm bare}|Z}/e$, the final value of $\lambda$ will lie
within observational bounds only if
\begin{equation}
e |\lambda_{\rm bare}|^{1/2} Z^{-1/2} < 10^{-120} \kappa_4^{-4}\ .
\end{equation}
Using the results above for $e$ and $Z$, the left-hand side (dropping
$2\pi$'s) is
\begin{equation}
|\lambda_{\rm bare}|^{1/2} \kappa_4^{1/3} V_7^{-1/3}
\sim |\lambda_{\rm bare}|^{1/2} \kappa_4 M_{11}^3
  \ .
\end{equation}
The step size is minimized in the low-energy string scenario
\cite{ArkDim98}, where $\lambda_{\rm bare}$ and $M_{11}$ are both
TeV-scale, but even in this case it is far too large, $10^{-75}
\kappa_4^{-4}$.

This is the `gap problem': the Brown-Teitelboim mechanism requires an
energy spacing which is infinitesimal compared to the scales of
microphysics.  In the next subsection we will consider
compactification with multiple four-forms, which can reduce the step
size to an acceptable value.

Because the compactification volume $V_7$ is a dynamical quantity and
not a fixed parameter, the four-form energy density is not a constant
but a potential for $V_7$.  In a realistic compactification this must
be stabilized, and the energetics of the four-form fluxes will enter
into the stabilization.\footnote {See for example the
discussions~\cite{Sun98,ArkDim98b}.}  Thus the volume $V_7$ itself
depends on $n$, and so the effective cosmological constant has
additional $n$-dependence beyond that included above.  For convenience
we will in the rest of this paper ignore this effect, treating the
geometry as fixed.

It should be noted that the allowed flux actually depends additively
on the values of flat background gauge potentials\footnote{We thank
E.~Witten for pointing this out.} --- these are just stringy
generalizations of the Schwinger model $\theta$-parameter.  As these
backgrounds vary the flux can take arbitrary real values.  This does
not, however, restore the original continuously variable cosmological
constant, because these background potentials are moduli and not
parameters.  As with the compactification geometry, these background
moduli must eventually be stabilized and so the fluxes will in fact
take discrete values.

\subsection{Multiple four-forms}

\begin{figure}[htb!]
  \hspace{.18\textwidth} \vbox{\epsfxsize=.64\textwidth
  \epsfbox{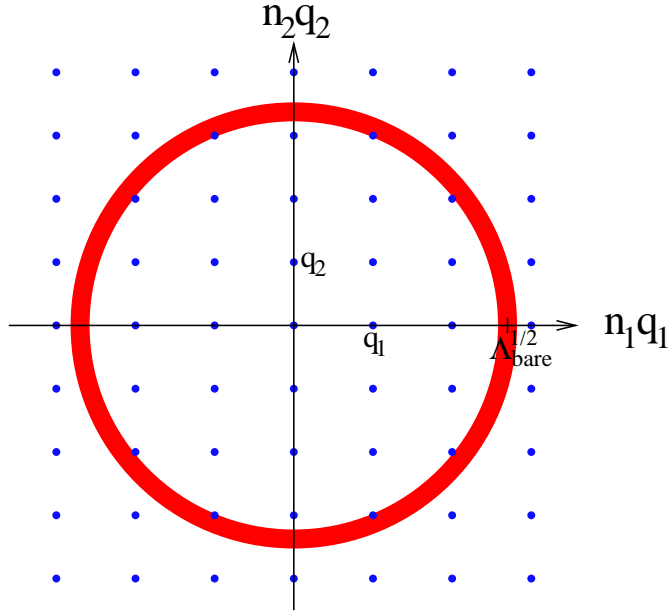}
}
  \caption[]%
  {\small\sl The allowed values of the four-form energy density are
  given by the radius-squared of points in the grid, whose dimension
  is the number of four-forms $J$.  The spacing in direction $i$ is
  $q_i$.  The negative of the bare cosmological constant corresponds
  to a $(J-1)$-dimensional sphere, and cancellation is possible if
  there is at least one grid point sufficiently close to the sphere.}
  \label{fig-grid}
\end{figure}

General compactifications actually give rise to several four-form fluxes,
and this can solve the gap problem.  Let there be $J$ such fluxes, with
\begin{equation}
\lambda = \lambda_{\rm bare} + \frac{1}{2} \sum_{i=1}^{J}
n_i^2 q_i^2 \label{eq-grid}\ .
\end{equation}
The question is whether there exists a set of $n_i$ such
that
\begin{equation}
2 |\lambda_{\rm bare}| <
\sum_{i=1}^{J} {n_i^2 q_i^2 } <
2(|\lambda_{\rm bare}| + \Delta\lambda)\ ,
\end{equation}
where $\Delta\lambda$ corresponds to the observational bound, roughly
$10^{-120}$ in Planck units.
This can be visualized in terms of a
$J$-dimensional grid of points, spaced by $q_i$ and labeled by $n_i$
(see Fig.~\ref{fig-grid}).
Consider a sphere of radius $r={|2 \lambda_{\rm bare}|}^{1/2}$
centered at $n_i=0$.  If one of the points $(n_1, n_2, \ldots, n_J)$
is sufficiently close to the sphere, the field configuration
corresponding to this point will lead to an acceptable value of the
cosmological constant.

More precisely, one should think of a thin shell, whose width encodes
the width of the observational range,
\begin{equation}
\Delta r = |2\lambda_{\rm bare}|^{-1/2} \Delta\lambda\ .
\end{equation}
We need at least one point to lie within the shell.  As we will
discuss, there may be large degeneracies --- let the typical
degeneracy be $D$. The volume per $D$ grid points must then be less
than the volume of the shell, $\omega_{J-1} r^{J-1} \Delta r$, where
the area of a unit sphere is $\omega_{J-1} = 2 \pi^{J/2} /
\Gamma(J/2)$.  Thus
\begin{equation}
\prod_{i=1}^J q_i \lesssim \frac{\omega_{J-1}}{D}|2 \lambda_{\rm
bare}|^{\frac{J}{2}-1} \Delta\lambda\ ,
\label{eq-condition}
\end{equation}
or
\begin{equation}
\frac{D}{\omega_{J-1}}
\prod_{i=1}^J \frac{q_i}{|2 \lambda_{\rm
bare}|^{\frac{1}{2}}}
\lesssim
\frac{\Delta\lambda_{}}{|2 \lambda_{\rm
bare}|}
\ .
\label{eq-condition2}
\end{equation}
In other words, the typical spacing of the spectrum of the
cosmological constant in a model with given $J$, $e_i$, and
$\lambda_{\rm bare}$ will be given by
\begin{equation}
{\Delta\lambda_{\rm min}} =
\frac{D \prod_{i=1}^J {q_i}}
{{\omega_{J-1}} |2 \lambda_{\rm
bare}|^{\frac{J}{2}-1}}\ .
\label{eq-spacing}
\end{equation}

An important feature of this result is that that the $q_i$ need not be
exceedingly small if there are more than two four-form fields.  In
order to achieve a small $\lambda$, it is sufficient that there be a
discrepancy between the magnitude of $\lambda_{\rm bare}$ and that of
the charges.  For fixed charges, the task of cancellation actually
becomes easier, the larger the bare cosmological constant.  This can
be understood from Fig.~\ref{fig-grid}.  The larger the shell, the
more points it will contain.%
\footnote{Note, however, that the radius of the shell in
Fig.~\ref{fig-grid} represents not $|2\lambda_{\rm bare}|$, but the
square root of $|2\lambda_{\rm bare}|$.  This is why one cannot
recognize in Fig.~\ref{fig-grid} the need for the charges $q_i$ to be
incommensurate, a fact that is immediately clear from
Eq.~(\ref{eq-grid}).  It is also the reason why increasing
$|\lambda_{\rm bare}|$ has no beneficial effect in the case of $J=2$.
For fixed $\Delta\lambda$, the shell gets thinner as one increases its
radius.  If $J=2$, this precisely compensates for the increase of the
shell radius, and the volume remains constant.} %
The results~(\ref{eq-condition}) to~(\ref{eq-spacing}) treat the $n_i$
as essentially continuous, and break down if any of the $q_i$ exceed
$J^{-1/2} |2 \lambda_{\rm bare}|^{1/2}$.  In this case the flux
associated with $q_i$ should simply be ignored.

For illustration suppose that $\lambda_{\rm bare}$ is at the Planck
scale, $(\sqrt 2\kappa_4)^4 \lambda_{\rm bare} \sim 1$, that the
number of four-forms is 100, a number which is large but not
unrealistic, and that $D$ is small. Then the
inequality~(\ref{eq-condition2}) implies that the typical charge must
be of order $10^{-1.6}$ in Planck units; note that $q_i$ is a
mass-squared, so we should perhaps measure the smallness by the square
root, $10^{-0.8} \sim 1/6$.  However, the assumption of no degeneracy
is rather optimistic, as we will discuss in the next subsection.

\subsection{M Theory Compactification}

Consider the compactification of M theory on a general manifold $K$.
The total number of fluxes is $J = N_3 + 1$, where $N_3$ is the number
of nontrivial three-cycles of $K$.  For each nontrivial three-cycle
$C_i$ there is a harmonic three-form $\mathbf{\omega}_{{\mathbf
3},i}$, and the seven-form field strength can be expanded
\begin{equation}
\mathbf{F_7} =
\frac{1}{M_{11}^3} \sum_{i=1}^{N_3} \mathbf{F}_{\mathbf{4},i}(x) \wedge
\mathbf{\omega}_{{\mathbf 3},i}(y)
+ *\mathbf{F}_{\mathbf{4},N_3+1}(x) \wedge \mathbf{\epsilon_7}
(y) \ . \label{eq-f7-f4}
\end{equation}
Here $\mathbf{\epsilon_7}$ is the volume form on $K$, so that
$\mathbf{F}_{\mathbf{4},N_3+1}$ is the flux discussed previously,
obtained simply by reduction of the eleven-dimensional flux.
Coordinates have been labeled as follows:
\begin{equation}
(X^0,\ldots,X^{11}) =
(x^0,\ldots,x^3,y^1,\ldots,y^7) \equiv (x^\mu,y^m)\ .
\end{equation}
Associated to each flux $\mathbf{F}_{\mathbf{4},i}$ is a
four-dimensional domain wall (membrane), obtained by wrapping three
legs of the M5-brane on $C_i$.

Let us illustrate this by a simple model, in which $K$ is simply a
seven-torus with flat internal metric $\delta_{mn}$, and with $y^m$
identified with period $2\pi r_m$; then $V_7 = \prod_{m=1}^7 (2\pi
r_m)$.  There is one three-cycle $C_i$ for each unordered triplet
$(m_i,m'_i,m''_i)$, or $( {7\atop 3} ) = 35$ in all.  The volume and
three-form associated with $C_i$ are
\begin{equation}
V_{3,i} = (2\pi)^3 r_{m_i} r_{m'_i} r_{m''_i}\ ,\quad
\mathbf{\omega}_{{\mathbf 3},i} = \frac{1}{V_{3,i}} \mathbf{dy}^{m_i}
\wedge \mathbf{dy}^{m'_i} \wedge \mathbf{dy}^{m''_i}\ .
\end{equation}
The four-dimensional action is
\begin{equation}
S = \int d^4\!x\, \sqrt{-g} \biggl( \frac{1}{2\kappa_4^2} R -
    \lambda_{\rm bare} -  \frac{1}{2\cdot4!} \sum_{i = 1}^{N_3+1}
{Z_i} F_{4,i}^2
\biggr) + S_{\rm branes}\ .
\end{equation}
Here
\begin{equation}
Z_i = \frac{2\pi M_{11}^3 V_7 }{ V_{3,i}^2}\ \ (i \leq N_3)\ ,
\quad \frac{1}{2\kappa_4^2} = Z_{N_3+1} ={2\pi M_{11}^9 V_7 }\ .
\label{eq-k4}
\end{equation}
The bare cosmological constant has been added by hand in this model.
In a real compactification, negative energy density can arise from
positive scalar curvature or an orientifold plane, for example.  The
tension of a membrane wrapped on $C_i$ is
\begin{equation}
\tau_i = {2\pi M_{11}^6
V_{3,i}}\ \ (i \leq N_3)\ ,\quad \tau_{N_3+1} =
{2\pi M_{11}^3}\ . \label{eq-tau}
\end{equation}
Its coupling to the $j$'th three-form potential is
 \begin{equation}
e_{i,j} = e \delta_{ij} \ , \quad e = {2\pi M_{11}^3} \ .
\end{equation}
The quantization condition is
\begin{equation}
\mathbf{F}_{\mathbf{0},i} = \frac{e n_i}{Z_i}\ ,
\end{equation}
and the effective cosmological constant is
\begin{equation}
\lambda = \lambda_{\rm bare} + \sum_{i=1}^{N_3+1}
\frac{e^2 n_i^2}{2 Z_i} \ ,
\end{equation}
so that
\begin{equation}
q_i =
e Z_i^{-1/2}\ .
\end{equation}
Thus,
\begin{equation}
q_i = \frac{(2\pi)^{1/2} M_{11}^{3/2} V_{3,i}}{ V_7^{1/2} }\ \
(i \leq N_3)\ ,
\quad  q_{N_3+1} =\frac{(2\pi)^{1/2}}{ M_{11}^{3/2} V_7^{1/2} }\ .
\label{eq-e_i}
\end{equation}
Note that $q_i^2 = 2\kappa_4^2 \tau_i^2$ for all $i$.

If the radii are appropriately incommensurate then so are the charges.
However, the degeneracy $D$ is still nontrivial, $2^J$, from $n_i \to
-n_i$ for each $i$ (note that in the $J=100$ model this reduces
$q_i^{1/2}$, but only by $\sqrt 2$).  This can be reduced to $D=2$ by
skewing the torus, which couples the different $n_i$.  However, if the
stabilization respects the symmetries of the torus there will be an
even larger degeneracy: permutations of the axes, obviously, and much
more --- the full $E_{7(7)}$ $U$-duality~\cite{HulTow95}.  The
resulting $D$ could significantly change the density of levels.  A
less symmetric compactification will have a much smaller duality
group, but we do not know how to estimate a reasonable degeneracy.
This effect becomes less important with fewer fluxes.

\subsection{Small charges from large dimensions}

From Eq.~(\ref{eq-condition}) one finds that a Planck-size
cosmological constant can be cancelled in a model with $j = 100$ types
of membranes with $q_i^{1/2}$ of order $1/6$ in Planck units.  If only
a few four-forms are present, much smaller charges will be required.
For examples, with $j=6$ and $D$ small one would need $q_i^{1/2} \sim
10^{-10} $ in Planck units.  However, these are not small quantities
compared to other numbers in elementary particle physics.  Indeed,
small membrane charges can be related to the gauge hierarchy, if the
latter arises by confining the gauge fields to a three-brane living in
eleven dimensions and taking some of the extra dimensions to be large,
as proposed in Ref.~\cite{ArkDim98}.

The large internal dimensions will play a double role here.  They are
the origin of the gauge hierarchy.  But in addition, they will lead to
small membrane charges, if membranes arise by wrapping a five-brane as
described in the previous subsection.  This amounts to reducing the
gauge hierarchy problem and the cosmological constant problem to the
single problem of stabilizing large radii.

In such models the fundamental scale $M_{11} $ is assumed to be near a
TeV.  The reduction~$(2\kappa_4^2)^{-1} = 2\pi M_{11}^9 V_7$ then
determines $V_7$ to be large in fundamental units.  For illustration,
consider again the seven-torus, with $k$ large dimensions of size
\begin{equation}
2\pi r_l = \frac{1}{M_{11}} ( V_7 M_{11}^7 )^{1/k},~~
l=1, \ldots, k\ ,
\label{eq-large-dims}
\end{equation}
and $7-k$ dimensions of radius 1 in fundamental units:
\begin{equation}
2\pi r_l = \frac{1}{M_{11}},~~ l=k+1, \ldots, 7\ .
\label{eq-v3}
\end{equation}
(In general, of course, the radii could have a range of different
sizes.  It is trivial to extend this discussion accordingly.)
Recalling the charges
\begin{equation}
q_i = \frac{(2\pi)^{1/2} M_{11}^{3/2} V_{3,i}}{V_7^{1/2}}\ \ (i<J)\ ,
\quad  q_{J} =\frac{(2\pi)^{1/2} }{ M_{11}^{3/2} V_7^{1/2} }\ ,
\label{eq-qs}
\end{equation}
it is most favorable to consider only $q_J$ plus those $q_i$ for which
all the dimensions are small.  For these, of which there are $ J_{0} =
( { 7-k \atop 3 }) $ the charges $q_i = q_J$.

We will consider a more general compactification with the same number
and sizes of dimensions, but not be restricted by the $J_0$ attainable
on the torus.  The condition~(\ref{eq-condition}) that the charges
$q_i$ allow for a sufficiently dense spectrum for $\lambda$ becomes
\begin{equation}
\frac{\omega_{J'-1}}{D}|2\lambda_{\rm
bare}|^{\frac{J'}{2}-1}
\Delta\lambda
\gtrsim \prod_{i=1}^J q_i
= (2 \pi)^{J'/2} M_{11}^{-3J'/2} V_7^{-J'/2}
= (2 \pi)^{J'/2} M_{11}^{3J'} \kappa_4^{J'}
\ ,
\label{eq-condition-2}
\end{equation}
where $J' \equiv J_0+1$.

What is the bare cosmological constant in models with large extra
dimensions?  It receives contributions from the tension of the
three-brane, $\lambda_{\rm brane}$, and from the bulk vacuum energy,
$\lambda_{\rm bulk}$ (as usual, all contributions from quantum field
theory are taken to be subsumed in these quantities)~\cite{Sun98,ArkDim98b}:
\begin{equation}
\lambda_{\rm bare} =
\lambda_{\rm brane} + V_7 \lambda_{\rm bulk}\ .
\label{eq-mm-bare}
\end{equation}
The most natural value for the brane tension is
\begin{equation}
\lambda_{\rm brane}
\sim 2 \pi M_{11}^4\ .
\label{eq-lower-lbare}
\end{equation}
(This value does not follow uniquely from the fundamental theory.  The
factor of $2\pi$ has been included to mimic the form of the M2- and
M5-brane tensions.)
It is natural (but not necessary) to assume that $\lambda_{\rm bulk}$
is generated by supersymmetry breaking on the brane.  This suppresses
the vacuum energy by a factor of the compact volume: $\lambda_{\rm
bulk} \sim 2 \pi M_{11}^4 / V_7$, so that both terms in
Eq.~(\ref{eq-mm-bare}) will be of order $2 \pi M_{11}^4$.  (Indeed,
the cosmological constant problem in these models amounts to the
assumption that the two terms cancel---an assumption that obviously
will not be made here.)  Supersymmetry breaking in the bulk could lead
to a higher value for $|\lambda_{\rm bulk}|$; ultimately, the only
constraint comes from bulk stability~\cite{ArkDim98b}, which is
weaker.  Recall however that the cancellation mechanism becomes more
accurate, the {\em larger} the magnitude of $\lambda_{\rm bare}$.  We
can therefore work with the value of Eq.~(\ref{eq-lower-lbare}).

The condition on the charges now becomes
\begin{equation}
(2^{-1/2} \kappa_4 M_{11})^{J'+4} \lesssim
10^{-120}\frac{\omega_{J'-1}}{\pi D} \ .
\end{equation}
For the extreme low-dimension picture, where $M_{11}$ is of order 1
TeV, this allows the very modest value $J' = 4$, independent of the
number $k$ of large dimensions.  This assumes that $D$ is not
enormous, as is reasonable for a small value of $J'$.  If we increase
$J'$ to 5 then $M_{11}$ can increase to 30 TeV.

If we take the value $\kappa_4 M_{11} \sim 10^{-1.5}$ that is
appropriate to the Witten GUT scenario~\cite{Wit96}, then we need a
large number of fluxes, again of order 100 (the precise number is
sensitive to uncertain numerical factors, for example in $\lambda_{\rm
bare}$).  Note also that this requires a cosmological constant of
order the GUT scale; a weak-scale cosmological constant cannot be
cancelled by our mechanism in this case.

\sect{Cosmology}
\label{sec-cosmology}

In the previous section we showed that multiple four-form strengths
arise in most M theory compactifications, and that these could lead to
a spectrum of effective cosmological constants sufficiently finely
spaced that some would lie in the observational range.  We must now
ask why the cosmological constant that we see actually takes such a
small value.

\subsection{The Brown-Teitelboim mechanism}
\label{sec-bt}

There are two possible approaches.  One could attempt to use the
framework of quantum cosmology to argue that the universe was created
with $\lambda$ equal to the smallest positive value in the
spectrum~\cite{Haw84b,Duf89,DunJen90}.  The other possibility is to
identify a dynamical mechanism by which an appropriate value of
$\lambda$ is obtained.

We will not follow the quantum cosmology approach.  It has the
disadvantage that the creation of a space-time from nothing (as
opposed to the quantum creation of objects on a given background) is
not well understood and possibly ill-defined.  The wave-function of
Hartle and Hawking~\cite{HarHaw83} would indeed be sharply peaked at
the smallest possible value for the cosmological constant.  But this
would include the effective cosmological constant from any inflaton
potential $V(\phi)$, so that there would not be any period of
inflation in generic models.  The proposals of Linde~\cite{Lin84b} and
Vilenkin~\cite{Vil86}, on the other hand, would give preference to a
large effective cosmological constant, which could come from any
combination of contributions from the four-forms and the inflaton.  To
cancel the cosmological constant one would then need a dynamical
effect anyway.

Thus we will employ a dynamical mechanism based on the creation of
membranes.  This is the approach followed by Brown and Teitelboim
(BT)~\cite{BroTei87,BroTei88}, who considered the first model
discussed in Sec.~\ref{sec-fourform}, with a bare cosmological
constant and a single four-form field strength.  We will review the
dynamics of this case before we generalize the mechanism to multiple
four-forms.

BT take $\lambda_{\rm bare}$ to be negative and $n$ large and
positive, so that $\lambda>0$.  Thus the universe will initially be
described by de~Sitter space.  On this background, membrane bubbles
can nucleate spontaneously.  They appear at a critical size and then
expand.  This is a non-perturbative quantum effect.  Its
semi-classical amplitude can be estimated from the Euclidean action of
appropriate instanton solutions~\cite{BroTei87,BroTei88}.  Inside the
membrane, the value of $n$ will be lower or higher by 1, and
correspondingly the cosmological constant changes by $(\pm n + 1/2)
q^2$.

Increase of $n$ occurs though a dominantly gravitational instanton.
It has no equivalent in non-compact spaces, as follows immediately
from energy conservation.  The instanton for a decreasing cosmological
constant is similar to the Coleman instanton for false vacuum decay in
flat space~\cite{Col77}, with a small correction from gravity.
Consequently, the amplitude for increasing the cosmological constant
is vastly more suppressed than that for decrease, and one may neglect
increase.

Starting from a generic, large value of $\lambda$, repeated membrane
creation thus produces de~Sitter regions with smaller and smaller
cosmological constant.  The nucleation rate decreases with $\lambda$.
For a certain range of parameters, membrane creation becomes
infinitely suppressed by gravitational effects~\cite{ColDel80} once
$\lambda$ is no longer positive.

The BT process is analogous to the neutralization of an electric field
(an $\mathbf{F_2}$) wrapped around a circle in a (1+1)-dimensional
world.  The Schwinger pair creation of (zero-dimensional) charged
particles decreases the field until it has too little energy to
nucleate another pair.

BT identified two problems with this scenario.  One is the `empty
universe problem', which we will address in Sec.~\ref{sec-empty}.
The other is the `gap problem', that in order for some values of the
cosmological constant to lie within the observational window one needs
membrane charges that are enormously small compared to the ordinary
scales of microphysics.  We have shown that this problem may be absent
in a theory with multiple four-forms.

Let us therefore extend the BT mechanism to the case of $J$
four-forms.  In the simplest models one can take $n_i \geq 0$ without
loss of generality.  For the initial configuration
$(n_{1,\rm{initial}}, \ldots, n_{J,\rm{initial}})$ we only need to
assume that the corresponding cosmological constant is positive:
\begin{equation}
\lambda_{\rm bare} + \frac{1}{2} \sum_{i=1}^J
n_{i,\rm{initial}}^2 q_i^2 > 0.
\end{equation}
This condition is generic, in the sense that it excludes only a finite
number of configurations.  In particular, if the unification scale is
high and $J \sim 100$, the charges $q_i$ can be large ($\sim 10^{-1}$)
and the inequality will be satisfied with $n_{i,\rm{initial}} \sim
1$.  If the unification scale is lower, the initial fluxes must be
greater.  But in such models large fluxes are often needed in any case
to stabilize the internal dimensions.

It will be convenient to assume the slightly stronger initial condition
that
$n_{i,\rm{initial}} \geq n_{i,\rm{obs}}$ for all $i$, where
$(n_{1,\rm{obs}}, \ldots, n_{J,\rm{obs}})$ denotes some
configuration that lies in the observational window.  This will permit
us to neglect the strongly gravitational instantons responsible for
increasing the $n_i$.

On the initial de~Sitter background, $J$ different types of membranes
can be nucleated through appropriate BT instantons.  Inside a membrane
of the $i$'th type, the flux $n_i$ is lowered by 1, and the
cosmological constant is lowered by $(n_i - 1/2) q_i^2$.  Although
membranes expand at the speed of light, they typically never
collide~\cite{GutWei83}, because they are embedded in de~Sitter space
and cannot catch up with its expansion.  Thus the ambient de~Sitter
space perdures eternally, harboring all types of membranes for which
$n_i>0$.  The same applies iteratively to the de~Sitter regions with
lower cosmological constant within each bubble.  Thus, all
combinations ($n_1, \ldots, n_J$) with $n_i \leq n_{i,\rm{initial}}$
and $\lambda > 0$ are attained, including those with $\lambda$ in the
observational range.

In the grid picture, Fig.~\ref{fig-grid}, the initial configuration
corresponds to a grid point some distance outside the sphere, in the
$n_i>0$ quadrant.  When a membrane of the $i$'th type is nucleated,
the configuration in its interior corresponds to the neighboring grid
point in the negative $n_i$ direction.  Nested membranes correspond to
a random walk in the grid.  Since each membrane bubble harbors all
other types of membranes (at least as long as $\lambda >0$), all such
paths through the grid are realized in the universe.  Overall, the
membrane dynamics corresponds to diffusion through the grid.  Every
point is populated via many different paths.

\subsection{The empty universe problem}
\label{sec-empty}

\subsubsection{Inflation and reheating}

The BT process involves spontaneous membrane nucleation in a prolonged
de~Sitter phase.  One would expect this to lead to an empty universe.
Particles are produced when a slow-roll field reaches a minimum of its
potential and starts to oscillate.  By this process, known as
reheating, slow-roll inflation avoids the empty universe problem of
old inflation.  Because membrane nucleation is highly suppressed, it
takes an exponentially long time to attain a suitable flux
configuration.  All fields will reach their vacua long before.
Particles may be produced, but they will be wiped away by the
remaining phase of the de~Sitter expansion.  In the end, it appears,
we may have achieved too much of nothing: a (nearly) vanishing
cosmological constant, but also vanishing entropy.

We will now discuss how this problem may be resolved.  If any
slow-roll field exists at all, one can argue that the problem does not
occur in multiple four-form models with unification at the GUT scale
or above, because the high temperature of de~Sitter space before the
final membrane nucleation kicks the inflaton out of its minimum.
Moreover, if the inflaton potential contains a false vacuum, inflation
and reheating can also occur in multiple four-form models with low
unification scale.

With an inflaton field included, the effective cosmological constant
is given by
\begin{equation}
\lambda_{\rm eff}(\phi) = \lambda_{\rm bare} + \frac{1}{2}
 \sum_{i=1}^{J} {n_i^2 q_i^2 } + V(\phi) \ .
\end{equation}
We take the inflaton potential $V(\phi)$ to have a stable minimum at
$\phi=0$.  By absorption into $\lambda_{\rm bare}$ we can arrange
$V(0)=0$.  The criterion for a suitable configuration $(n_1, \ldots,
n_J)$ is that the cosmological constant be small for $\phi=0$:
$\lambda_{\rm eff}(0) = \lambda \approx 0$.  For $\phi \neq 0$ one
obtains a positive effective cosmological constant, $\lambda_{\rm
eff}(\phi) = V(\phi)$, at least temporarily.

Slow-roll inflation with $\lambda=0$ is described as follows.  An
inflaton field $\phi$ rolls down in a potential $V(\phi)$.  During
this time, the universe expands exponentially, like de~Sitter space
with an effective cosmological constant $\lambda_{\rm eff}(\phi) =
V(\phi)$.  Quantum fluctuations during this era freeze when they leave
the horizon, forming seeds for density perturbations.  When $\phi$
reaches the bottom of the potential, it oscillates, inflation ends,
and the universe is reheated.

The inflaton potential must be very flat.  We will not address the
difficult problem of how such potentials may arise from a fundamental
theory; we note merely that they must exist if inflation is the
correct explanation for homogeneity and density perturbations.  For
inflation to last long enough, one must require that the initial value
of the inflaton field, $\phi_0$, is sufficiently far from the minimum
of $V(\phi)$:
\begin{equation}
\phi_0 \geq \phi_* \ ,
\end{equation}
where $\phi_*$ corresponds to sixty $e$-foldings of de~Sitter-like
expansion.  This is realized, for example, if the inflaton field is
initially trapped in a false vacuum far enough from the true minimum
at $\phi=0$.  More generically, suitable domains will exist if one
assumes chaotic initial conditions in the early universe~\cite{Lin83}.
In the scenario we have described, however, one must worry that any
inflaton field will reach its minimum when $\lambda$ is still large,
because membrane creation takes an exponentially long time.
Consequently, $\phi$ would no longer be available to perturb and
reheat the universe when the flux configuration corresponding to
$\lambda \approx 0$ is reached.

\subsubsection{Kicking the inflaton}

The evolution of $\phi$ is actually a combination of classical
slow-roll and Brown\-ian motion~\cite{Vil83b,Lin86a}.  The latter can
be understood as a random walk induced by the Gibbons-Hawking
temperature~\cite{GibHaw77a} of de~Sitter space:
\begin{equation}
T(\phi) = \frac{H(\phi)}{2\pi} \ ,
\end{equation}
where the Hubble parameter is given by%
\footnote{We use the `reduced' Planck mass, $M_{\rm Pl} = (8 \pi
G_{\rm N})^{-1/2} = \kappa_4^{-1} = 2.43 \cdot 10^{18}$ GeV.} %
\begin{equation}
H(\phi)^2 = \frac{\lambda_{\rm eff}(\phi)}{3 M_{\rm Pl}^2}.
\end{equation}
The characteristic time scale in de~Sitter space is the Hubble time,
$\Delta t = H^{-1}$.  A typical quantum fluctuation of the field
$\phi$, during the time $\Delta t$, is given by~\cite{Lin82b}
\begin{equation}
|\delta \phi| = \sqrt{2} T(\phi) = \frac{H(\phi)}{\sqrt{2} \pi} \ .
\end{equation}

The classical decrease $|\Delta \phi|$ of the inflaton field can be
estimated from the restoring force, $-V'(\phi)$:
\begin{equation}
|\Delta \phi| \approx \frac{1}{2} V'(\phi) (\Delta t)^2 =
 \frac{V'(\phi)}{2H(\phi)^2} \ .
\end{equation}
A prime denotes differentiation with respect to $\phi$.  We neglect
velocity effects because they are small during slow-roll and average
to zero in Brown\-ian motion.  The random walk dominates over
classical evolution if $|\delta \phi| > |\Delta \phi|$, or
\begin{equation}
\sqrt{\frac{2}{3}} \frac{1}{3\pi}\,
\lambda_{\rm eff}(\phi)^{3/2} M_{\rm Pl}^{-3} > V'(\phi) \ .
\label{eq-pen}
\end{equation}

We are interested in the temperature of the universe just before a
final membrane is nucleated.  Consider a flux configuration $(n_1,
\ldots, n_{j-1}, n_j+1, n_{j+1}, \ldots, n_J)$, where $(n_1, \ldots,
n_{j-1}, n_j, n_{j+1}, \ldots, n_J)$ corresponds to a cosmological
constant in the observational window.  If the charges $q_i$ are large,
the penultimate cosmological constant,
\begin{equation}
\lambda_{\rm pen} = \left(n_j - \frac{1}{2} \right) q_j^2,
\end{equation}
will be large.  Since $\lambda_{\rm eff,pen}(\phi) \geq \lambda_{\rm
pen} > 0$, Eq.~(\ref{eq-pen}) will be satisfied for a range of values
of $\phi$ including $\phi=0$.  Therefore the inflaton will take random
values within a (finite or infinite) neighborhood of $\phi=0$.

When the final membrane is nucleated, the temperature in its interior
suddenly vanishes.  Then the inflaton no longer experiences
significant Brown\-ian motion and rolls to its minimum.  The question
is whether this period of inflation is sufficient.  (Less ambitiously,
one could ask only whether $\phi$ will be large enough to reheat the
universe.)  One would like the width of the random distribution of
$\phi$ to be a few times larger than $\phi_*$, the value required for
60 $e$-foldings.  Then it will be likely that $\phi \geq \phi_*$ at
the time of the final membrane nucleation.

From Eqs.~(\ref{eq-k4}), (\ref{eq-v3}), and (\ref{eq-qs}) one obtains
\begin{equation}
q_j^2 \approx 8 \pi^2 \left( \frac{M_{11}}{M_{\rm Pl}} \right)^6
M_{\rm Pl}^4.
\end{equation}
The inequality, Eq.~(\ref{eq-pen}), will be satisfied if
\begin{equation}
(2 n_j -1)^{3/2} \sqrt{\frac{2}{3}} \frac{8\pi^2}{3} \left(
\frac{M_{11}}{M_{\rm Pl}} \right)^9 M_{\rm Pl}^3 > V'(\phi).
\label{eq-ninth}
\end{equation}

One may take as examples the polynomial potentials $V(\phi) = 10^{-12}
M_{\rm Pl}^2 \phi^2/2$ and $V(\phi) = 10^{-14} \phi^4/4$, and note
that $\phi_* \approx 15 M_{\rm Pl}$ in both cases.  Then
Eq.~(\ref{eq-pen}) must be satisfied with $V' \gtrsim 10^{-10} M_{\rm
Pl}^3$ for sufficient inflation ($\phi>\phi_*$) to be likely.  The
condition becomes
\begin{equation}
\frac{M_{11}}{M_{\rm Pl}} \gtrsim (2 n_j - 1)^{-1/6} \cdot 10^{-1.3}.
\end{equation}
Thus the universe will undergo a normal period of inflation if the
unification scale is $10^{17}$ GeV or higher.

The $n_j$-dependent factor does not contribute much since $J$ is large
and the flux numbers will be of order one.  Because of the $ (
M_{11}/M_{\rm Pl} )^9$ suppression in Eq.~(\ref{eq-ninth}), this
mechanism rapidly becomes inefficient for lower unification scale.

\subsubsection{Trapping the inflaton}

There is an alternative approach to the empty universe problem.  It is
less generic, but has the advantage that it can work in models with
low unification scale, $M_{11} \geq 1$ TeV.  Assume that the potential
of the inflaton field (or of any other field with suitable coupling to
the inflaton) contains a false vacuum.  During the long de~Sitter era
before the $\lambda \approx 0$ flux configuration is attained, this
vacuum will of course decay by Coleman-De Lucia
tunnelling~\cite{ColDel80}.  As we discussed earlier, Guth and
Weinberg have shown that bubbles do not percolate in de~Sitter
space~\cite{GutWei83}.  Because the bubbles do not all collide, we
need not fear that the entire universe will be converted to the true
vacuum.  In the ambient metastable space, the BT mechanism proceeds as
before.  Eventually, it produces regions where $\lambda \approx 0$
while $\lambda_{\rm eff}$ is given by the energy density of the false
vacuum.  Only then will we be interested in the decay of the false
vacuum.  After tunnelling, the field emerges on the other side of the
barrier.  For a wide class of potentials, the field configuration at
this point will still be far from the true vacuum.  The fields can
then roll to the minimum, thus inflating and reheating the universe.

The understanding of the cosmology of models with large internal
dimensions is still being developed.  This makes a detailed
implementation of the generalized BT mechanism difficult.  We should
caution that there are important constraints on inflationary models
that operate after radius stabilization~\cite{KalLin98}.

In this subsection it has been assumed that the effective potential
$V(\phi)$ is the same before and after the final membrane nucleation.
However, there can be important corrections from the high temperature
before the final transition~\cite{DinRan95}.  They will typically make
the potential steeper but also shift its minima.  Thus, after the
final nucleation, the inflaton field will not be in a local minimum
and can roll down.  Moreover, one would expect coupling constants of
the effective field theory to depend on the fluxes.  This also
contributes to the flux-dependence of the effective potential and
generically to a shift of its minima when a membrane is nucleated.%
\footnote{We thank L.~Susskind and S.~Thomas for pointing this out to
us.}

\subsection{Vacuum selection}

In the BT scenario the universe generically develops a large number of
different, exponentially large regions with every value of the
cosmological constant in the discretuum, including large values.  Why
are we located in one of the regions with a small cosmological
constant?

Most regions will not contain structure such as galaxies.  Observers
are necessarily located where structure does form, which restricts us
to regions in the {\em Weinberg window\/}~\cite{Wei87},
\begin{equation}
-10^{-120} M_{\rm Pl}^4 < \lambda < 10^{-118} M_{\rm Pl}^4.
\label{eq-weinberg}
\end{equation}
The upper bound is about 100 times larger than the observed $\lambda$.
It is obtained by demanding that the cosmological constant must not
dominate the evolution of the universe before a redshift of about 4,
so that gravitational clustering operates long enough for galaxies to
form.  The lower bound follows because the universe must not
recollapse while stars and galaxies form.  Its magnitude is comparable
to the observed cosmological constant, but it has opposite sign.  Much
work has been devoted to strengthening these constraints by more
careful astrophysical and statistical arguments (see, e.g.,
\cite{Efs95,Wei96,MarSha97,GarVil99} and references therein).

Such considerations may be distasteful to some but should not be
viewed as an easy fix.  They cannot be applied unless the fundamental
theory satisfies a number of rather non-trivial conditions: it must
admit different values of the cosmological constant; they must contain
at least one value in the observational range; and there must be a
dynamical mechanism that allows some regions to attain such a value.
The aim of this paper has been to present evidence that all of these
conditions may be satisfied in compactified 11D supergravity.

\subsection{Stability}

In order for our picture to be satisfactory we need the rate of bubble
nucleation from a phase with small cosmological constant to be small
on the scale of the age of our universe.  The tunnelling amplitude is
proportional to $e^{-B}$, where $B$ is the normalized action of the
corresponding instanton~\cite{BroTei87,BroTei88}.  A sufficient
condition is $B \gtrsim 10^3$.

We consider a single membrane, which changes the flux $j$ from $n_j$
to $n_j -1$.  The domain wall tension is $\tau_j$, given in
Eq. (\ref{eq-tau}), and the change in the cosmological constant is
\begin{equation}
\delta\lambda = -\left(n_j - \frac{1}{2}\right) q_j^2 = -2 M_{\rm
Pl}^{-2} \left(n_j-\frac{1}{2}\right) \tau_j^2\ .
\end{equation}
For $n_j \gg 1$, gravity has negligible effects and the action is
given by
\begin{equation}
B = \frac{27\pi^2}{2 \left(n_j - \frac{1}{2}\right)^3 (2
M_{\rm Pl}^{-2} q_j)^2} \ .
\end{equation}
To estimate the $n_i$ we assume approximate equipartition of the
energy among the fluxes so that
\begin{equation}
\frac{n_i^2 q_i^2}{2} \approx \frac{2\pi M_{11}^4}{J}\ .
\label{eq-part}
\end{equation}
For the nonzero fluxes in section~2.6, $\tau_i = 2\pi M_{11}^3$, and
$q_i^2 = 8 \pi^2 (M_{11}/M_{\rm Pl})^2 M_{11}^4$.  We obtain
\begin{equation}
B \approx \frac{27 \pi^{3/2} J^{3/2}}{16 \sqrt{2} (M_{11}/M_{\rm
Pl})^3}\ .
\end{equation}
In the large dimension case $B$ is of order $10^{46}$ and so the
tunneling is negligible.  Even for the Witten GUT scenario, where $J
\sim 100$, it is of order $10^8$ and again tunneling is negligible.

At higher unification scales, for which $M_{11}/M_{\rm Pl}>10^{-1.5}$,
one finds $J > 100$.  Then Eq.~(\ref{eq-part}) yields $n_i<1$ and thus
breaks down.  Almost all relevant configurations will have $n_i \in
\{0,1\}$ for all $i \in \{1, \ldots, J\}$.  We can therefore assume
$n_j = 1$.  In this case the additional suppression due to gravity is
significant, although it is never total for our parameters.  One finds
\begin{equation}
B = \frac{1728 \pi^2}{(2 M_{\rm Pl}^{-2} q_j)^2}
  = 54 ( M_{11}/M_{\rm Pl} )^{-6}\ .
\end{equation}
Tunnelling will be negligible for $M_{11}/M_{\rm Pl} < 0.6$.  Therefore
vacuum stability is not a significant constraint on our mechanism.  A
stronger constraint on $ M_{11}/M_{\rm Pl} $ is obtained from
Eq.~(\ref{eq-condition-2}) by requiring a realistic number of
three-cycles, say $J<10^3$.

\sect{Conclusions}
\label{sec-summary}

Compactifications of M-theory generally give rise to multiple
four-form field strengths.  We showed that such theories have vacua
with discrete but closely spaced values for the cosmological constant.
In the Witten GUT scenario, the spectrum will contain values of
$\lambda$ in the observable range if the number of four-forms is of
order 100.  (This requires that the cosmological constant to be
cancelled is of GUT scale, not weak scale).  In models with large
internal dimensions, four or five four-forms suffice, and a weak-scale
cosmological constant can be cancelled.  By repeated membrane
nucleation, flux configurations with $\lambda \approx 0$ arise
dynamically from generic initial conditions.  We argued that entropy
and density perturbations can be generated in such regions, and showed
that the amplitude for the decay of the $\lambda \approx 0$ vacuum is
negligible.

An attractive feature of this proposal is that it simultaneously
addresses two questions that are usually treated as distinct.  The
first question is: Why is the cosmological constant not huge?  One
would expect a vacuum density $\lambda$ of order $M_{\rm Pl}^4$, or at
least TeV$^4$ with supersymmetry.  Until recently this was the only
cosmological constant problem.  It appeared to require a symmetry
ensuring the exact cancellation of all contributions to the
cosmological constant.  This is difficult because contributions are
expected to come from many different scales.  The second question is:
Why is the cosmological constant not zero?  Recent evidence%
\footnote{A review of these observations can be found in
Ref.~\cite{Car00}.}
points to a flat universe with $\Omega_{\rm m} \approx 0.3$ and
$\Omega_\lambda \approx 0.7$.  The favored value for the vacuum energy
is $\lambda \approx 10^{-120} M_{\rm Pl}^4 \approx (0.003 \mbox{
  eV})^4$.  In particular, a flat universe with vanishing vacuum
energy has been ruled out.  But if it is difficult to explain
$\lambda=0$, a small non-zero cosmological constant seems to pose an
even greater theoretical challenge.  The mechanism we propose has
limited accuracy because of flux quantization, so that a residual
cosmological constant is inevitable.

Our proposal has certain features of the Brown-Teitelboim idea, and
also certain features of eternal inflation~\cite{Lin86a}.  Previously,
however, both of these ideas have been difficult to realize with a
plausible microphysics.  Our proposal allows both to be realized
within string theory.  For the Brown-Teitelboim idea, the main problem
was the very small energy scale needed in the discretuum; we see that
this can be obtained from a normal hierarchy with multiple fluxes.
Eternal inflation with generic polynomial potentials requires scalar
field expectation values strictly larger than the Planck scale.  In
string theory the scale of the field manifold is the string scale,
which is no larger than the Planck scale.  The manifold is actually
noncompact, but the asymptotic regions generally correspond to
decompactification of spacetime, and in this region the effective
potential generally ceases to be flat.  We have realized a version of
eternal inflation that does not require such a large scalar, and uses
elements already present in string theory.%
\footnote{A precursor to the idea of four-form-driven eternal
inflation was presented in Ref.~\cite{BouCha98}.}
Moreover, if the membrane charges are large, the high temperature of
de~Sitter space before the final membrane nucleation induces Brownian
motion of the inflaton field, thus preparing suitable initial
conditions for chaotic inflation after the transition.

The main problem with realizing our picture is the stabilization of
the compact dimensions, which is of course a ubiquitous problem in
string theory.  A positive bulk cosmological constant is a useful
ingredient~\cite{Sun98,ArkDim98b}, but it is not clear that this can
be realized in string theory.

It is interesting that the naked singularity
proposal~\cite{Ark00,Kac00} appears to lead in the end to a very
similar picture.  The free parameters that correspond to boundary
conditions at a naked singularity in a compact space will become, in a
four-dimensional effective Lagrangian, variable coupling constants.
In the original proposal these were assumed to be continuous and
constant in time, but in Ref.~\cite{PolStr00} it was argued that they
are discrete and can change across a domain wall, just as for the
fluxes considered here.  In the example~\cite{PolStr00} there was a
potentially large number of states, of order $e^{\sqrt N}$ where $N$
is at Ramond-Ramond charge of the singularity.  Note, however, that a
charge of order $10^5$ is needed to produce a discretuum sufficiently
dense to account for the smallness of the cosmological constant.  In
Ref.~\cite{PolStr00} the main focus was on supersymmetric states,
which were all degenerate, but with supersymmetry breaking there will
again be a spectrum for $\lambda$.  Again, stabilization will be an
issue.

The appearance of the anthropic principle, even in the weak form
encountered here, is not entirely pleasant, but we would argue that it
is necessary in any approach where the cosmological constant is a
dynamical variable.  That is, a small value for the present
cosmological constant cannot be obtained by dynamical considerations
alone.  The point is that we can follow cosmology at least back to
nucleosynthesis, when the present cosmological constant contributed
only a fraction $10^{-30}$ to the energy density of the universe, and
so was dynamically irrelevant.  At earlier times, including the point
where the cosmological constant is to have been determined, the
fraction would have been even smaller.\footnote{One exception is the
wormhole idea~\cite{Col88a}, where the value of the cosmological
constant in our universe is determined by the presence of other,
empty, universes.  At least one of the authors retains a certain wary
fondness for this possibility.}

\section*{Acknowledgments}

We would like to thank N.~Arkani-Hamed, S.~Dimopoulos, S.~Kachru,
A.~Lin\-de, J.~March-Russell, J.~Rahmfeld, L.~Susskind, S.~Thomas,
N.~Toumbas, and E.~Witten for useful discussions.  We would also like
to thank J.~Feng, J.~March-Russell, S.~Sethi, and F.~Wilczek for
informing us about their forthcoming work.  While completing this work
we learned that J.~Donoghue was also pursuing the idea of anthropic
determination of the cosmological constant with nondynamical
four-forms~\cite{Don00}.  The work of J.P.\ was supported by National
Science Foundation grants PHY94-07194 and PHY97-22022.  The work of
R.B.\ was supported by a BASF fellowship of the German National
Scholarship Foundation.

\bibliographystyle{board}
\bibliography{all}

\end{document}